\documentclass[twocolumn]{revtex4-1}
\usepackage{amsmath}
\usepackage{physics}
\usepackage{subcaption}
\usepackage{graphicx}
\usepackage{float}

\graphicspath{
{"/Users/mullspace/Dropbox (MIT)/currentcourses/6.443/project/allfiles/"}}

\begin{document}
\title{Two particle quantum walks with topological phases}
\author{Samuel Huberman}
\email{schuberm@mit.edu}
\affiliation{%
Mechanical Engineering, MIT
}%

\begin{abstract} 
The split step quantum walk for two noninteracting particles is numerically simulated. The entropy of entanglement and spatial particle distributions are calculated for a range of initial states and for a range of disorder. The impact of varying the topological phase on these quantities is discussed.

\end{abstract}

\maketitle
\section{Introduction}
Quantum walks lie at the intersection of physics and computation. For instance, universal quantum computation can be achieved through a quantum walk framework \cite{childs2009universal}. Extensions to searching for marked vertices of graphs and hypercubes with a quadratic speed up over the classical algorithm has been shown \cite{ambainis2003quantum}. There has also been work on applying quantum walks to the graph isomorphism problem \cite{berry2011two}.

Meanwhile, the concept of topology has recently emerged as exciting perspective in condensed matter physics. Since the discovery of the quantum Hall Effect and the subsequent explanation through the invocation of topological quantities \cite{PhysRevB.31.3372}, a rich field of study has emerged. The principle recipe of study has been to examine simple models with specific symmetries and calculate various topological invariants and dynamics as those symmetries are broken. For instance the Haldane model shows that as time reversal symmetry is broken, conductance will be quantized \cite{PhysRevLett.61.2015}. It is not surprising that one can apply this approach to the study of quantum walks.

The organization of this work is as follows; first a brief review of quantum walks is presented, followed by some comments on the current approach to studying topological phases in random walks. A framework for studying two particle quantum walks with topological phases is outlined. For different initial states (i.e. separable and entangled) and ranges of disorder, two particle distribution functions and the entropy of entanglement are numerically evaluated. Finally, some general comments and observations are made and unanswered questions are raised.

\section{Review of single particle quantum walks}

The most elementary of discrete quantum walks is the one-dimensional walk. This walk consists of ``coin'' operator in the spin Hilbert space, for instance, the Hadamard coin can be used
\begin{eqnarray}
C &= \frac{1}{\sqrt(2)}\begin{bmatrix}
    1    & 1  \\
     1 & -1      \\
\end{bmatrix}
\end{eqnarray}

and a translation operator in the space Hilbert space
\begin{eqnarray}
T &= \sum_x [ \ket{x+1}\bra{x} \otimes \ket{0}\bra{0} + \ket{x-1}\bra{x} \otimes \ket{1}\bra{1}] \\
  & = T_0 +T_1
\end{eqnarray}

which accordingly shifts the spin component $\ket{0}$ to $\ket{x+1}$ from $x$ and the spin component $\ket{1}$ to $\ket{x-1}$ from $x$. These operators are then multiplied to generate the unitary transformation
\begin{eqnarray}
U &=  T (I \otimes C)
\end{eqnarray}

Sequential application of $U$, yields a state such that after $N$ time steps
\begin{eqnarray}
\ket{\psi_N}&=  U^N \ket{\psi_0}
\end{eqnarray}

\begin{figure*}
\medskip
\begin{subfigure}{0.48\textwidth}
\includegraphics[width=\linewidth]{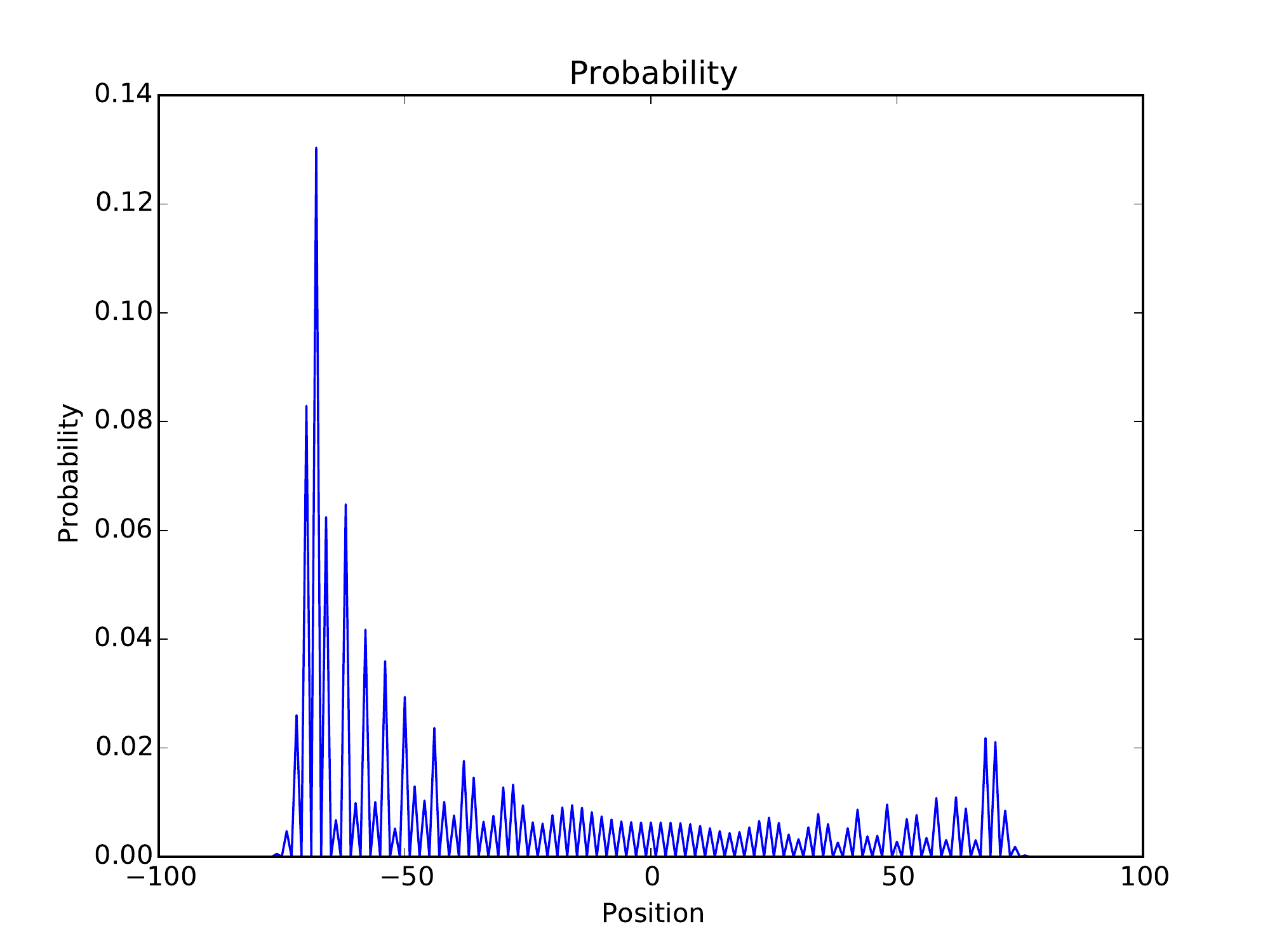}
\caption{Probability distribution after 100 time steps for the Hadamard walk.} \label{fig:1a}
\end{subfigure}\hspace*{\fill}
\begin{subfigure}{0.48\textwidth}
\includegraphics[width=\linewidth]{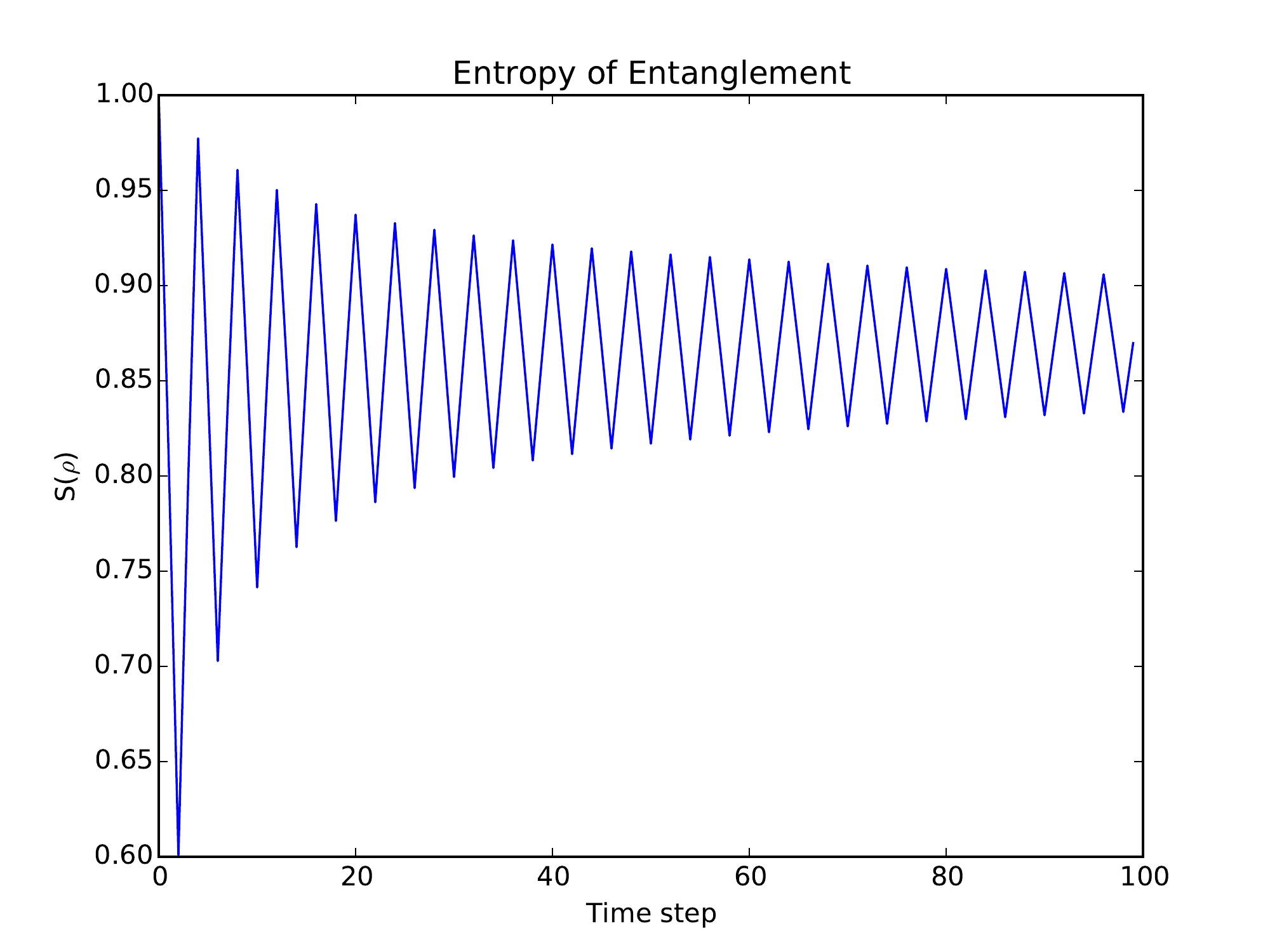}
\caption{Entropy as a function of time step for the Hadamard walk.} \label{fig:1b}
\end{subfigure}
\caption{Probability and Entropy for the one dimensional Hadamard walk. \label{fig:1}}
\end{figure*}

One of the appealing aspects of quantum walks is that the distribution spreads quadratically faster than the classical counterpart (i.e the time steps required to have a non zero amplitude at a given position for the first time, also known as the hitting time). In Figure~\ref{fig:1a}, the asymmetry of the probability distribution is observed which is a consequence of the chosen coin and the initial state. Symmetrical quantum walks can be constructed. In Figure~\ref{fig:1b}, the entropy of entanglement as a function of time can be seen to fluctuate and asymptote to a value 0.87, which is consistent with the value found in \cite{carneiro2005entanglement}. For a more in depth review of quantum walks, see \cite{venegas2012quantum, ambainis2003quantum}. 

\section{Topological quantum walks}

\begin{figure}[!h]
\begin{center}
\scalebox{1.5}{ \includegraphics{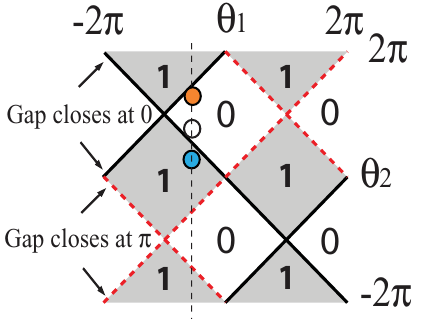}}
\caption{Topological Phase diagram for the split step quantum walk. Image taken from \cite{PhysRevA.82.033429}.}\label{fig:phase}
\end{center}
\end{figure}

What follows is a synopsis of the work done by Kitagawa et al. in \cite{PhysRevA.82.033429}. Taking inspiration from the Hamiltonians studied in other topological systems, Kitagawa et al. seek the appropriate parameterization of the unitary transformation to be applied in the quantum walk. First, the translation invariance of the translation operator enables the Fourier transformation to reciprocal space, where a Brillouin Zone can be defined. Secondly, the coin is chosen to be the rotation operator, such that the angle of rotation is free parameter. It was then shown that a topological invariant, $Z$, which is determined by the angle of rotation $\theta$, could be defined as the number of rotations the spin vector experiences as it moves through the Brillouin Zone. The number of rotations through the BZ will be either 0 or 1 and is mapped out in the phase diagram seen in Figure~\ref{fig:phase}. It was shown that the topological phases defined by this parameterization are equivalent to those of the Su-Schrieffer-Heeger model, a rich model for the study of topological phenomena. 

Kitagawa et al. further parameterize the Hamiltonian by incorporating a dependence on a pair of rotation angles (not necessary, but allows for a natural mapping of the phase diagram). The unitary transformation that encodes the evolution operator of this Hamiltonian is referred to as the split step quantum walk, defined by the following
\begin{equation}
U_{ss}(\theta_1,\theta_2) = T_1 R(\theta_2) T_0 R(\theta_1) 
\end{equation}

where $R(\theta)$ is the rotation operator in the spin Hilbert space
\begin{equation}
R(\theta) = 
\begin{bmatrix}
    \cos(\theta/2)    & -\sin(\theta/2)   \\
     \sin(\theta/2) & \cos(\theta/2)      \\
\end{bmatrix}
\end{equation}

and $T$ is the translation operator in the space Hilbert space
\begin{eqnarray}
T &= \sum_x [ \ket{x+1}\bra{x} \otimes \ket{0}\bra{0} + \ket{x-1}\bra{x} \otimes \ket{1}\bra{1}] \\
  & = T_0 +T_1.
\end{eqnarray}

As is done in the Hadamard quantum walk, the split step quantum walk evolved by repeatedly applying $U_{ss}$ to some initial state.

\section{Two particle quantum walks}
Two particle quantum walks have received little attention, partly because of the difficulty in analytically studying such systems. However, a few numerical studies exist \cite{berry2011two,annabestani2010asymptotic,rigovacca2015two}. This motivates the approach for this project in combination with the apparent lack of a published study of the two particle split quantum walk. For the two noninteracting particle quantum walk, we extend the Hilbert space of the split step unitary transformation
\begin{equation}
U_{ss,AB} = U_{ss,A}(\theta_1,\theta_2)  \otimes U_{ss,B} (\theta_1',\theta_2'),
\end{equation}
with the freedom that $\theta_1,\theta_2$ does not necessarily equal $\theta_1',\theta_2'$, enabling $Z_A$ and $Z_B$ to be distinct. The system in which $\theta_1,\theta_2$ is independent of $\theta_1',\theta_2'$ will be referred to as the Two Particle Two Phase Walk (TPTPW).

Typically, the tools to characterize two particle quantum walks are the joint probability distribution and the entanglement entropy \cite{berry2011two,annabestani2010asymptotic} and are put to use here, in hopes of finding some insight into the effects of the underlying topology. The joint probability distribution for finding particle $A$ in position $i$ and particle $B$ in position $j$ after $N$ steps is given by \cite{PhysRevA.74.042304}
\begin{multline}
P^{\pm}_{AB}(i,j;N) = \frac{1}{2}( P_{10}(i,j;N) + P_{01}(i,j;N)\\ \pm [ I_{10}(i;N)I_{01}(j;N) + I_{01}(i;N)I_{10}(j;N)])
\label{eq:prob}
\end{multline}
where
\begin{equation}
P_{10}(i,j;N) = P_0(i;N)P_1(j;N)
\end{equation}
and
\begin{equation}
I_{10}(i;N) = \bra{0,1}(U^\dagger)^N\ket{i}\bra{i}U^N\ket{0,0}.
\end{equation}

Note that Equation~\ref{eq:prob} is not a simple product of probabilities, but rather a sum containing information about the correlations due to the interference between the two states, that when linearly combined, form an entangled state. This can be seen when one takes an expectation of an observable given an entangled state and noting the cross terms in the complete expansion \cite{berry2011two}.

In order to characterize the entanglement between the two particles and their positions, we use the entropy of entanglement, which is given by
\begin{equation}
S(\rho_c) = -\Tr( \rho_c \log_2 \rho_c)
\end{equation}

where the reduced density matrix is obtained by tracing out the position degrees of freedom
\begin{equation}
\rho_c = \Tr_x( \rho_{cx}).
\end{equation}

Through the entropy of entanglement and the spatial probability distributions, the effects of disorder and initial states can be investigated. Three initial states will be examined: the entangled states $\ket{\psi^+_0}=  \frac{1}{\sqrt{2}}( \ket{01} +  \ket{10} )$ and $\ket{\psi^-_0}=  \frac{1}{\sqrt{2}}( \ket{01} -  \ket{10} )$ and the separable state $\ket{\psi_0}=   \ket{01} $. It should be noted that a general state, such as $\ket{\psi_0}  = ( \alpha\ket{01} +  \beta\ket{10} )$ can be used as is done in \cite{annabestani2010asymptotic}.

Disorder is introduced through the randomization of the $\theta_1,\theta_2$. In doing so, there is freedom in how disorder is implemented in the calculation. For instance, the $\theta_1,\theta_2$ for particle A can be randomized at each spatial point for each time step, while $\theta_1',\theta_2'$ for particle B remains fixed. This procedure is in contrast with other works, which implement random coins that are fixed in time \cite{konno2009one}. So while this procedure is not physically motivated, is a natural way to introduce the effect of randomness into the system. Within this text, weak disorder refers to a uniform distribution over the interval $[\theta-0.1\pi,\theta+0.1\pi]$, while strong disorder refers to a uniform distribution over the interval $[\theta-2\pi,\theta+2\pi]$.

Another tactic is to couple two distinct topological phases in the spatial domain, such that for instance, the $x < 0$ portion would belong to a specific topological phase while the $x > 0$ portion belongs to another, thus creating a topological phase boundary. This is implemented by selecting $\theta_1,\theta_2$ to be function of $x$. Calculations of entropy of entanglement and the spatial probability distributions for these systems, referred to as a Two Particle Topological Boundary Walk (TPTBW), were also performed. Note the distinction from the other system in studied here, the TPTPW.

\begin{figure*}
\medskip
\begin{subfigure}{0.4\textwidth}
\includegraphics[width=\linewidth]{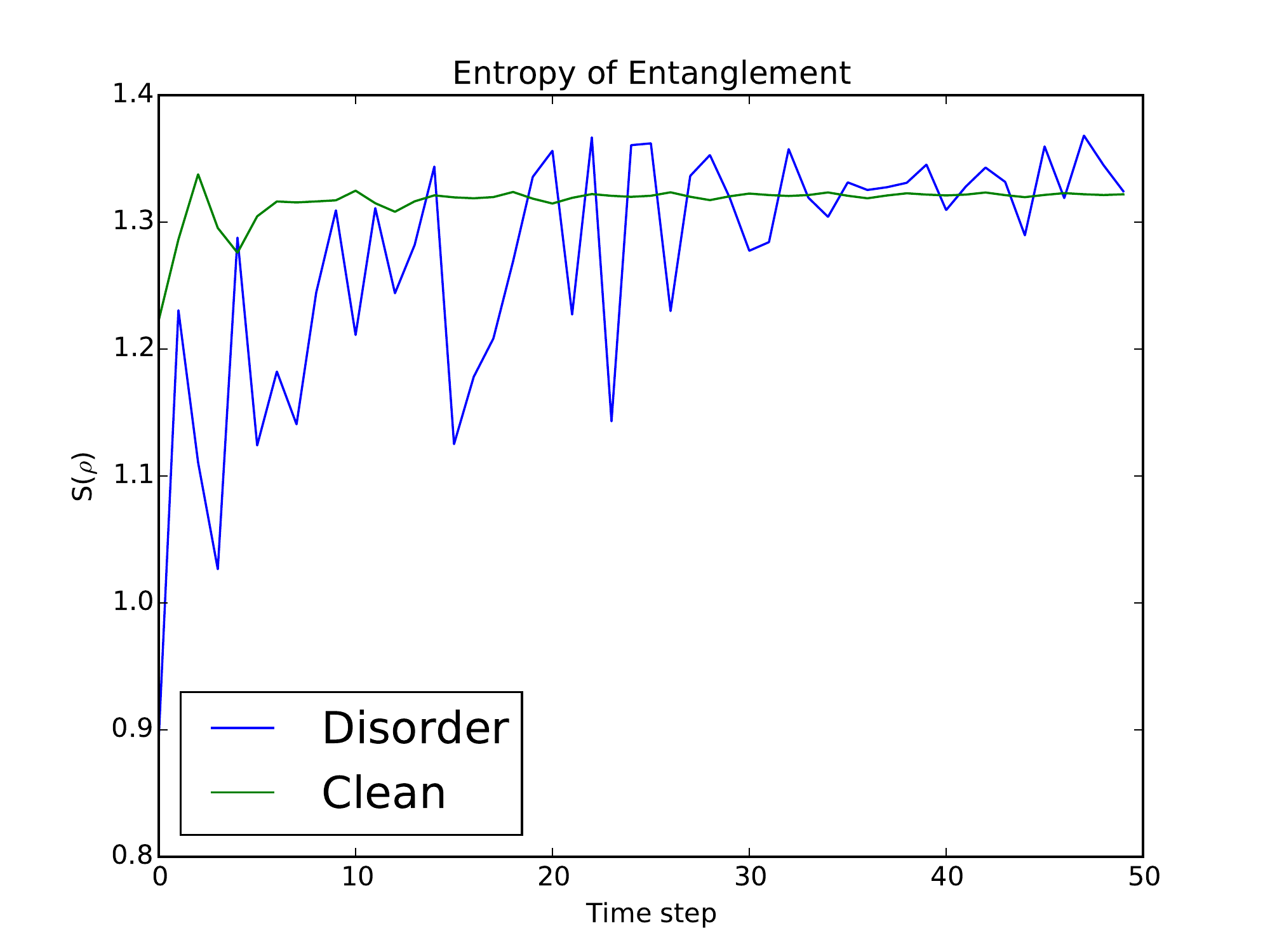}
\caption{Entropy as a function of time step for a TPTPW.}\label{fig:3a}
\end{subfigure}\hspace*{\fill}
\begin{subfigure}{0.4\textwidth}
\includegraphics[width=\linewidth]{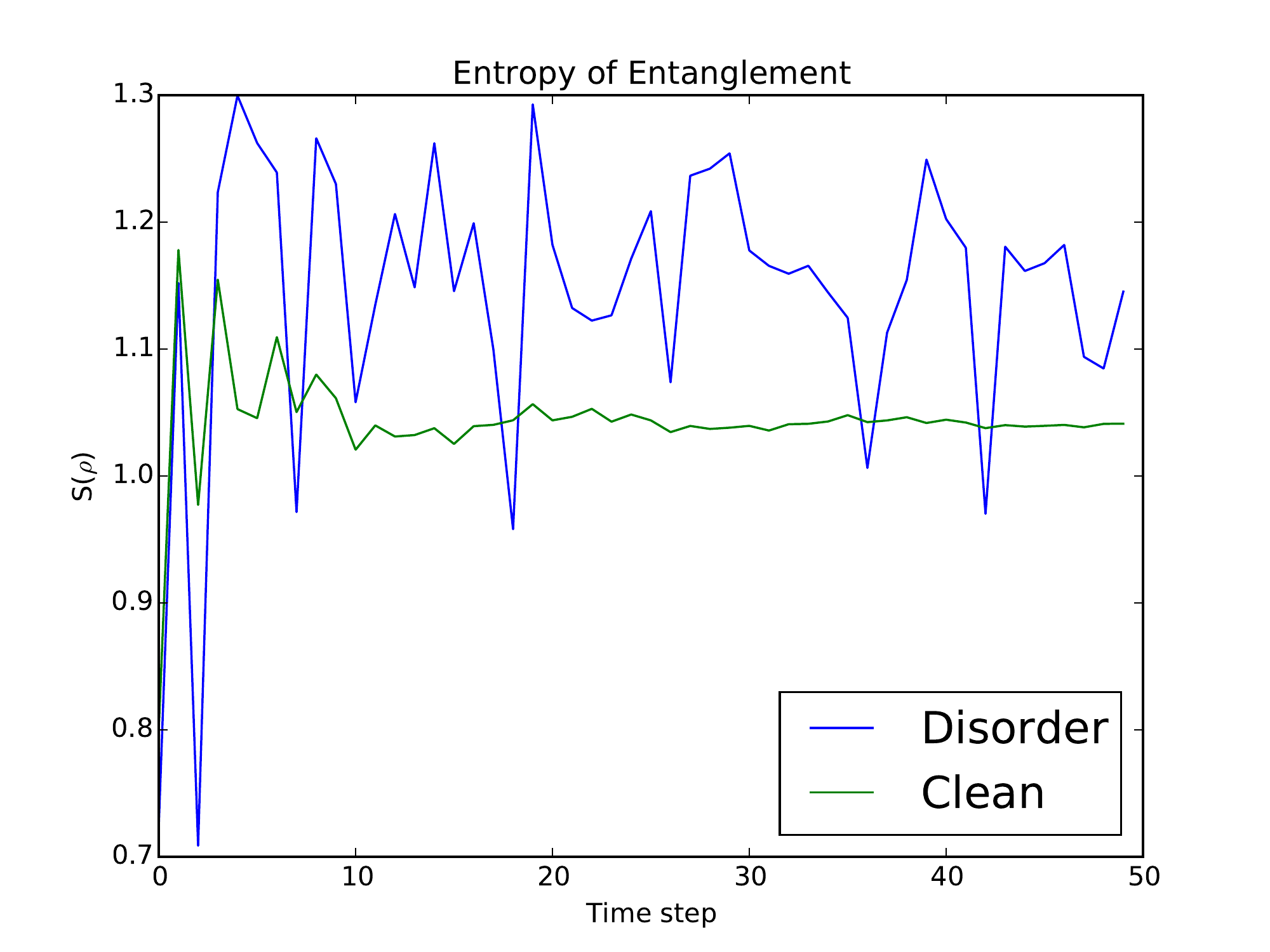}
\caption{Entropy as a function of time step for a TPTBW.}\label{fig:3b}
\end{subfigure}
\caption{Entropy for clean and strong disordered cases for the TPTPW and TPTBW}\label{fig:3}
\end{figure*}

\section{Results}

\begin{figure*}[ht] 
\begin{subfigure}{0.48\textwidth}
\includegraphics[width=\linewidth]{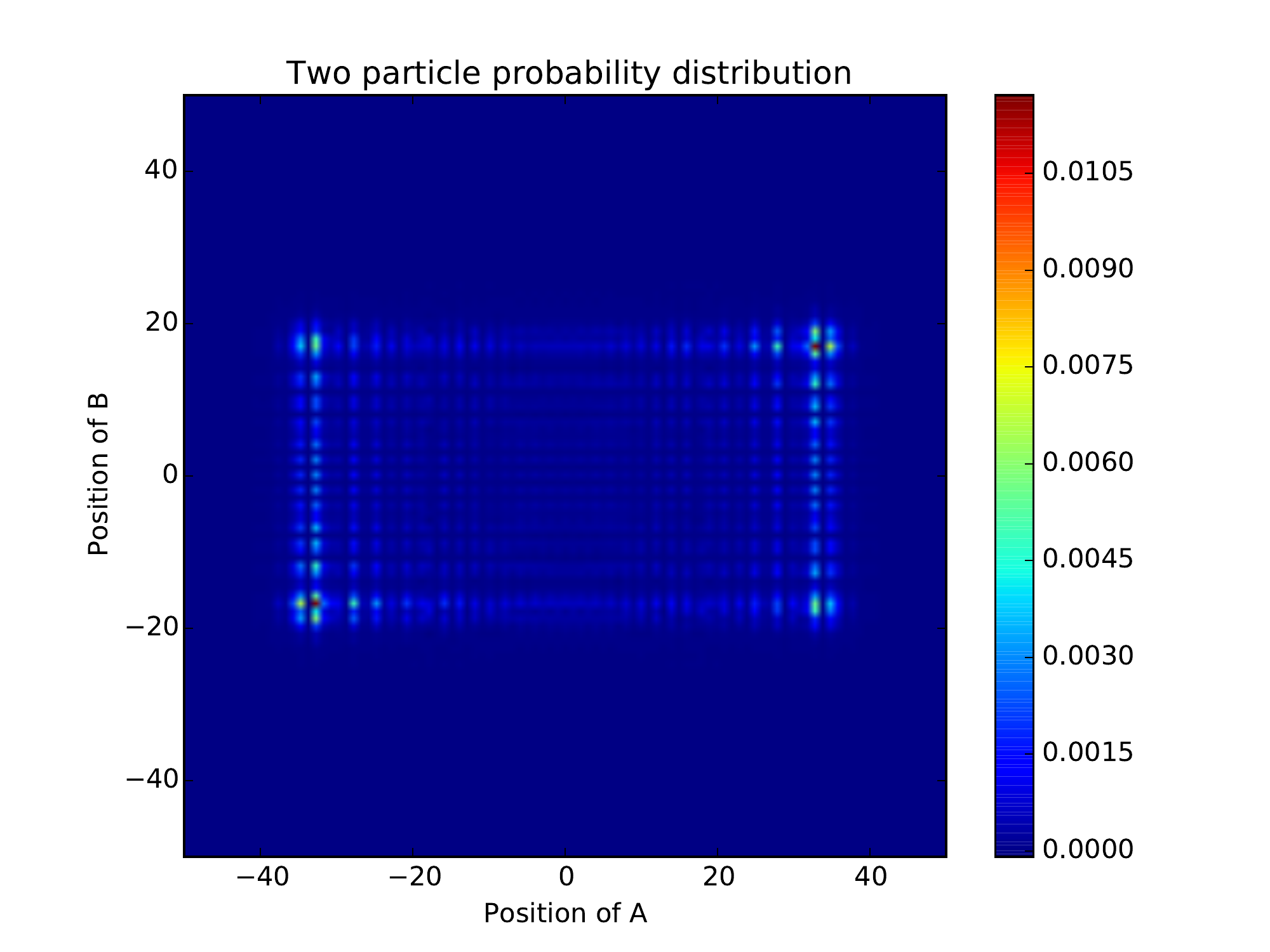}
\caption{TPTPW with $Z_A = 1, Z_B = 0$ and $\ket{\psi^+_0}$.}\label{fig:4a}
\end{subfigure}\hspace*{\fill}
\begin{subfigure}{0.48\textwidth}
\includegraphics[width=\linewidth]{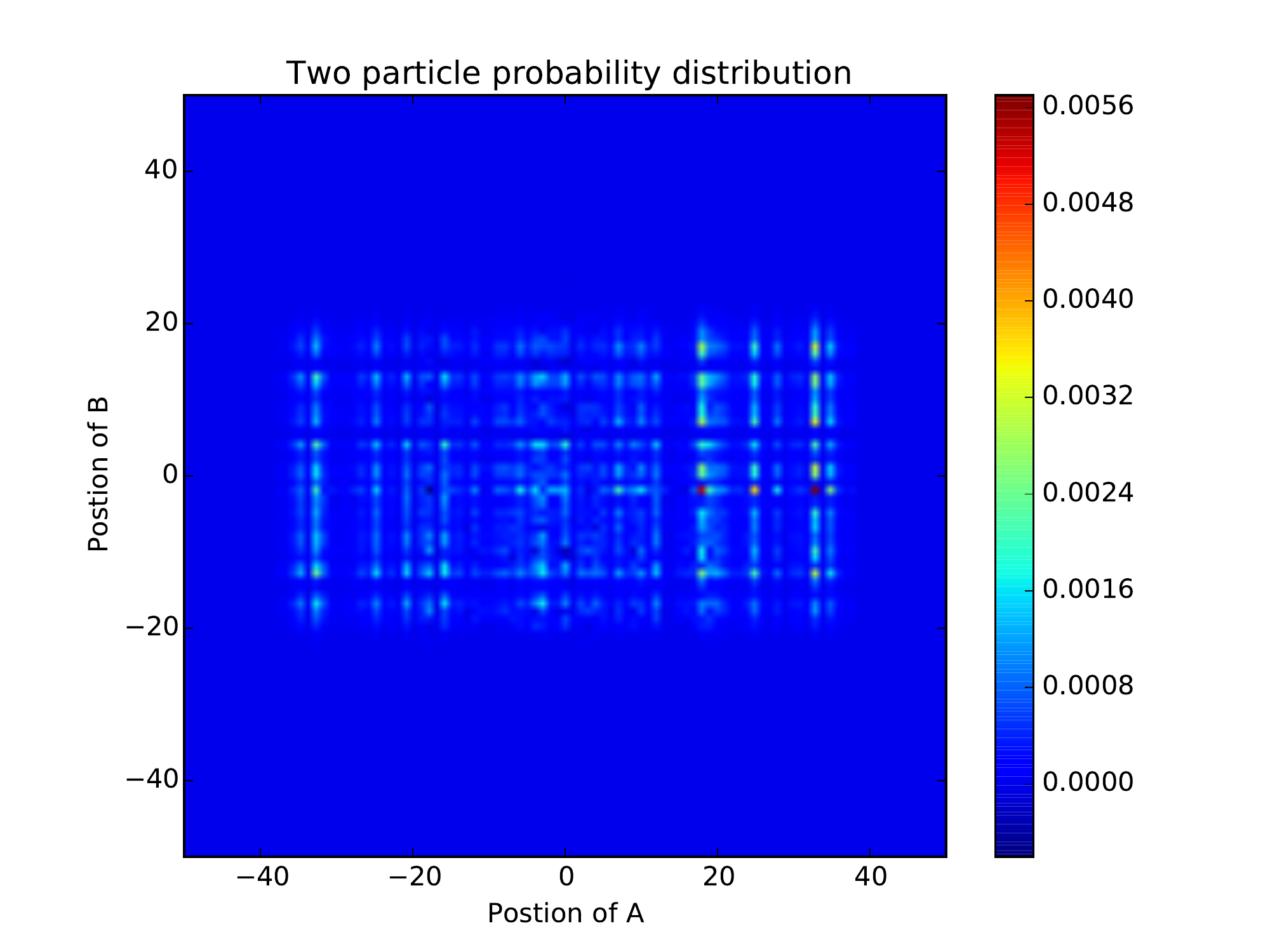}
\caption{TPTPW with $Z_A = 1, Z_B = 0$ and $\ket{\psi^+_0}$ with weak disorder.}\label{fig:4b}
\end{subfigure}

\medskip
\begin{subfigure}{0.48\textwidth}
\includegraphics[width=\linewidth]{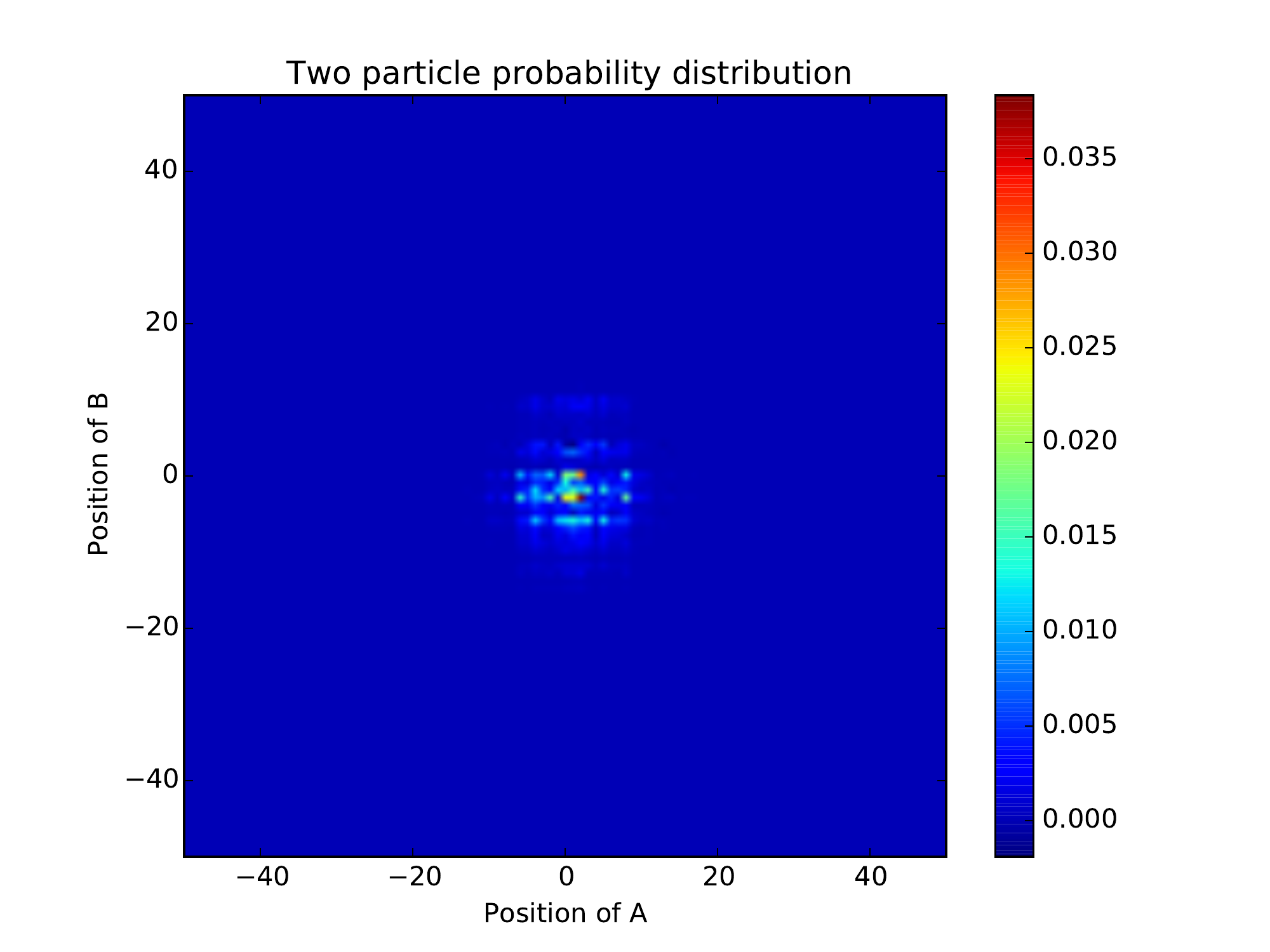}
\caption{TPTPW with $Z_A = 1, Z_B = 0$ and $\ket{\psi^+_0}$ with strong disorder.}\label{fig:4c}
\end{subfigure}\hspace*{\fill}
\begin{subfigure}{0.48\textwidth}
\includegraphics[width=\linewidth]{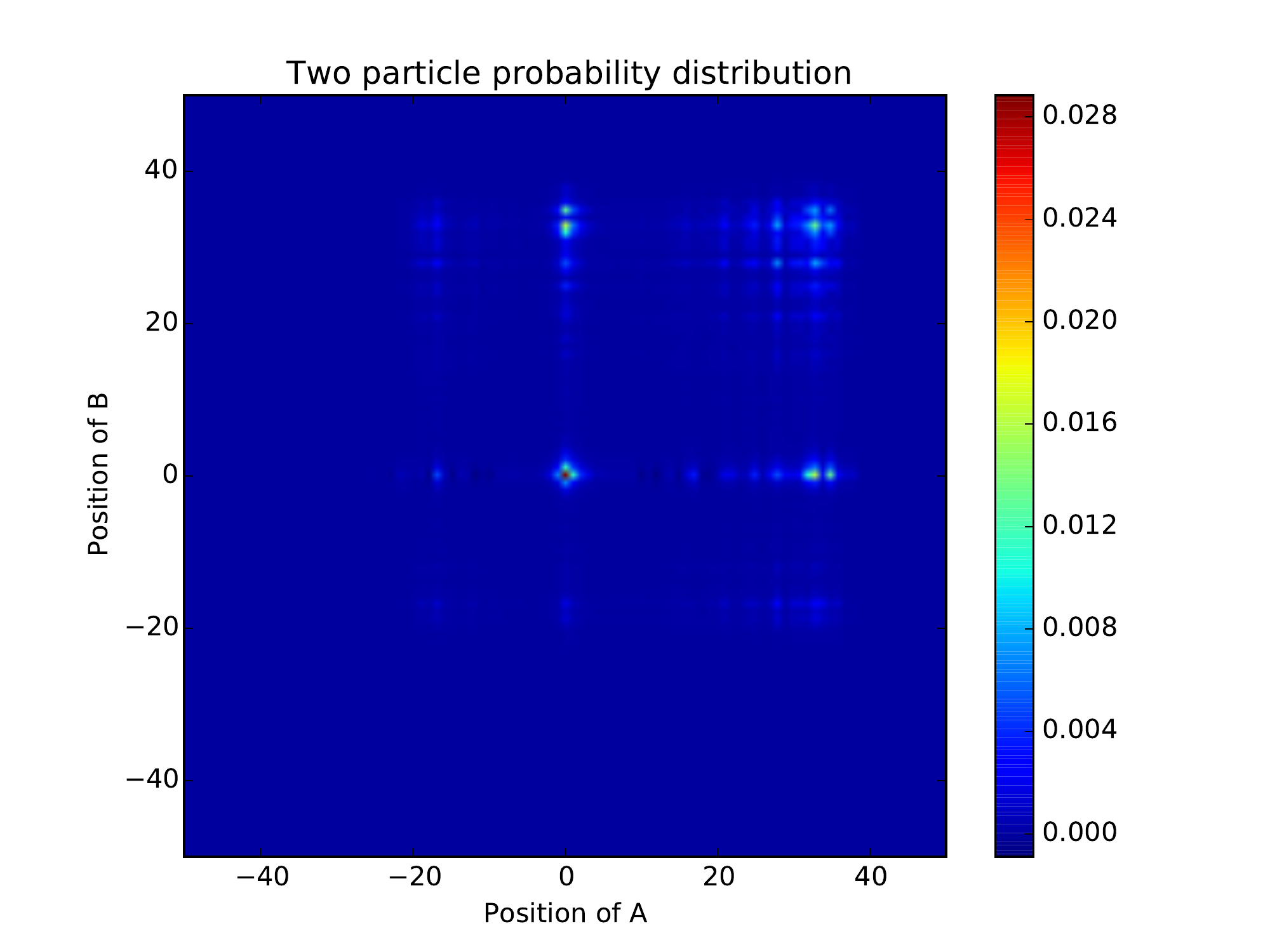}
\caption{TPTBW with $Z_{x^+} = 0$ and $Z_{x^-} = 1$.}\label{fig:4d}
\end{subfigure}

\medskip
\begin{subfigure}{0.48\textwidth}
\includegraphics[width=\linewidth]{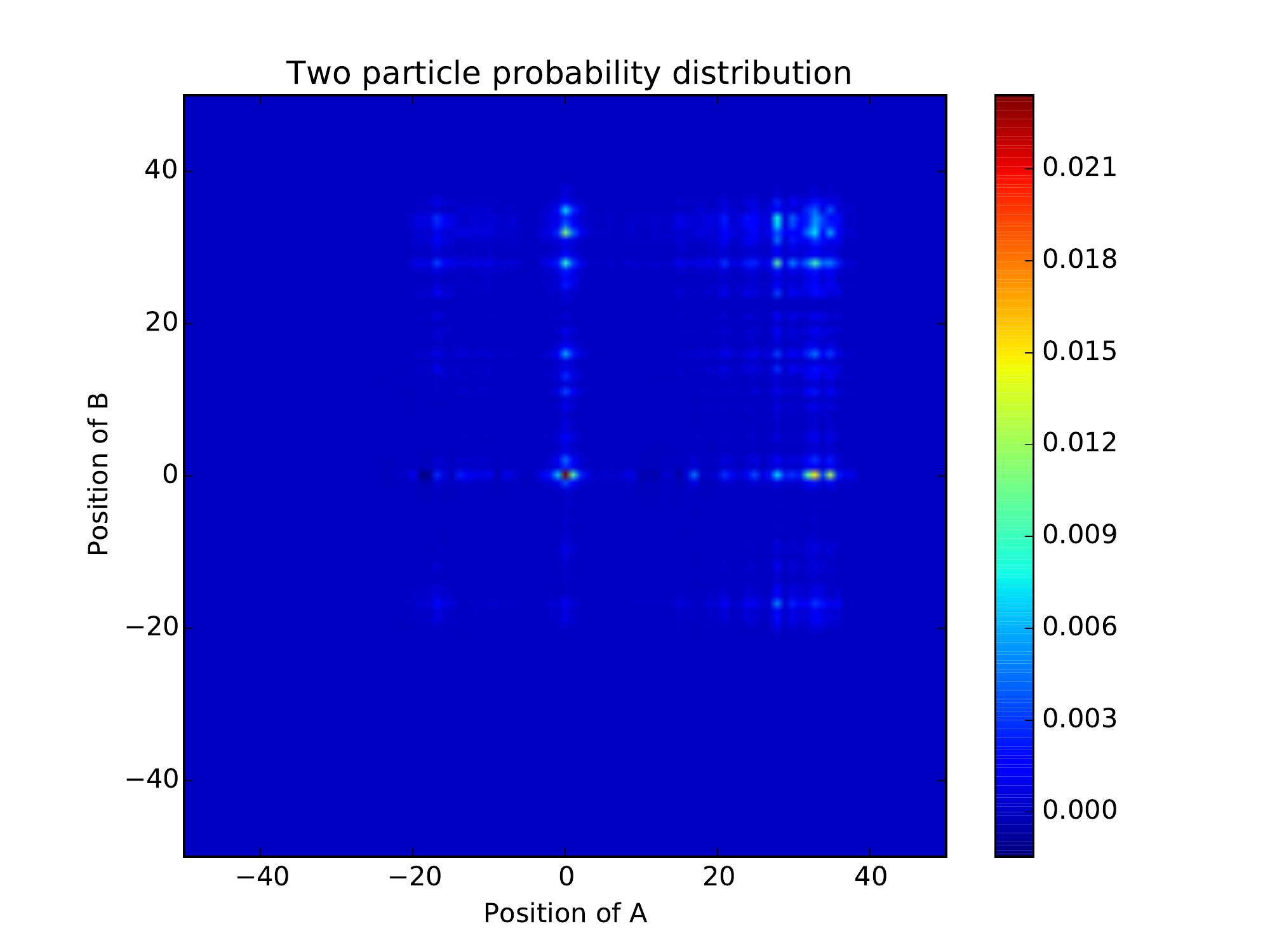}
\caption{TPTBW with $Z_{x^+} = 0$ and $Z_{x^-} = 1$ and $\ket{\psi^+_0}$ with weak disorder.}\label{fig:4e}
\end{subfigure}\hspace*{\fill}
\begin{subfigure}{0.48\textwidth}
\includegraphics[width=\linewidth]{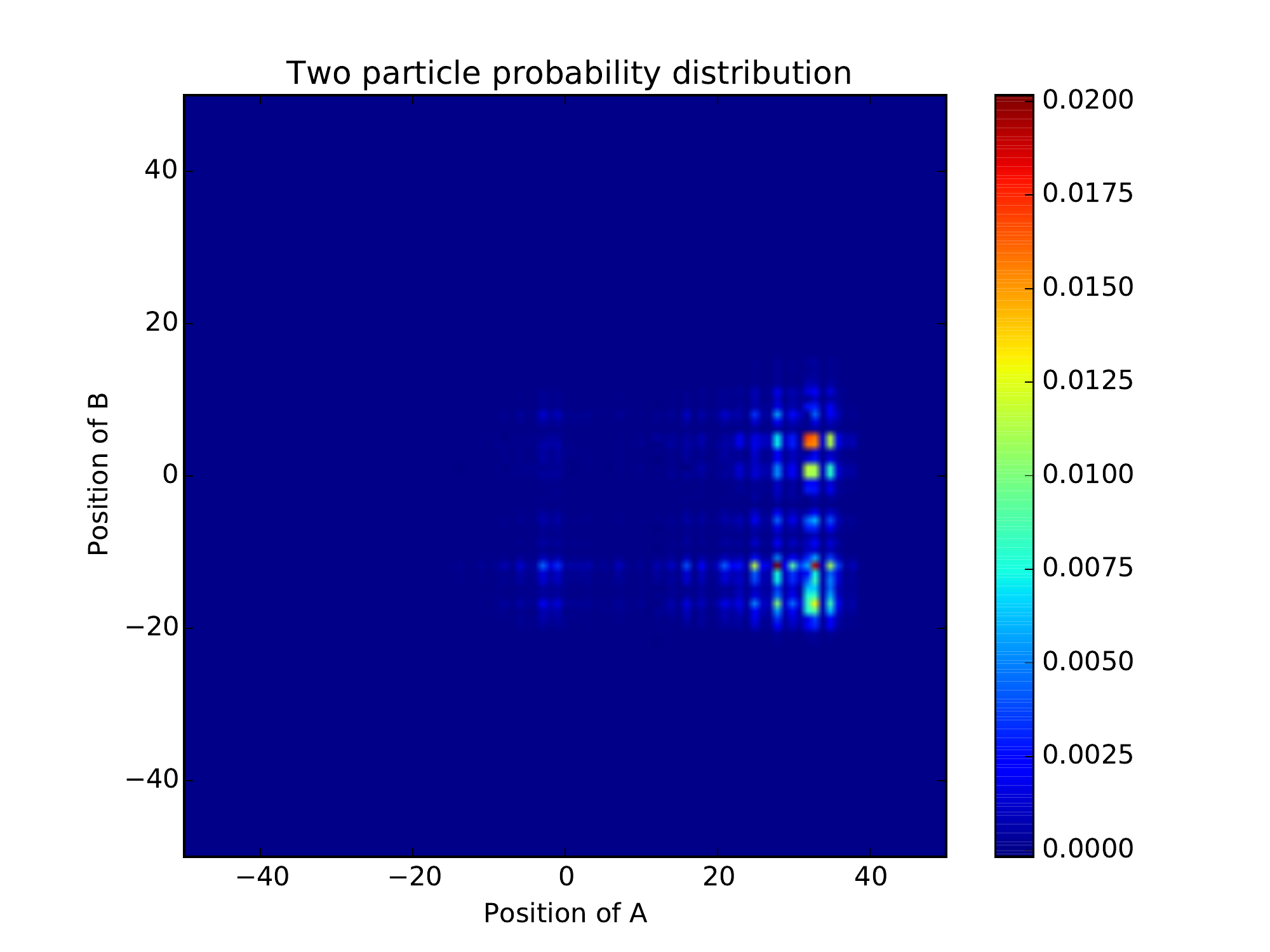}
\caption{TPTBW with $Z_{x^+} = 0$ and $Z_{x^-} = 1$ and $\ket{\psi^+_0}$ with strong disorder.}\label{fig:4f}
\end{subfigure}

\caption{Spatial two particle probability distributions for TPTPW and TPTBW.}\label{fig:4}
\end{figure*}

\begin{figure*}[ht!] 
\begin{subfigure}{0.48\textwidth}
\includegraphics[width=\linewidth]{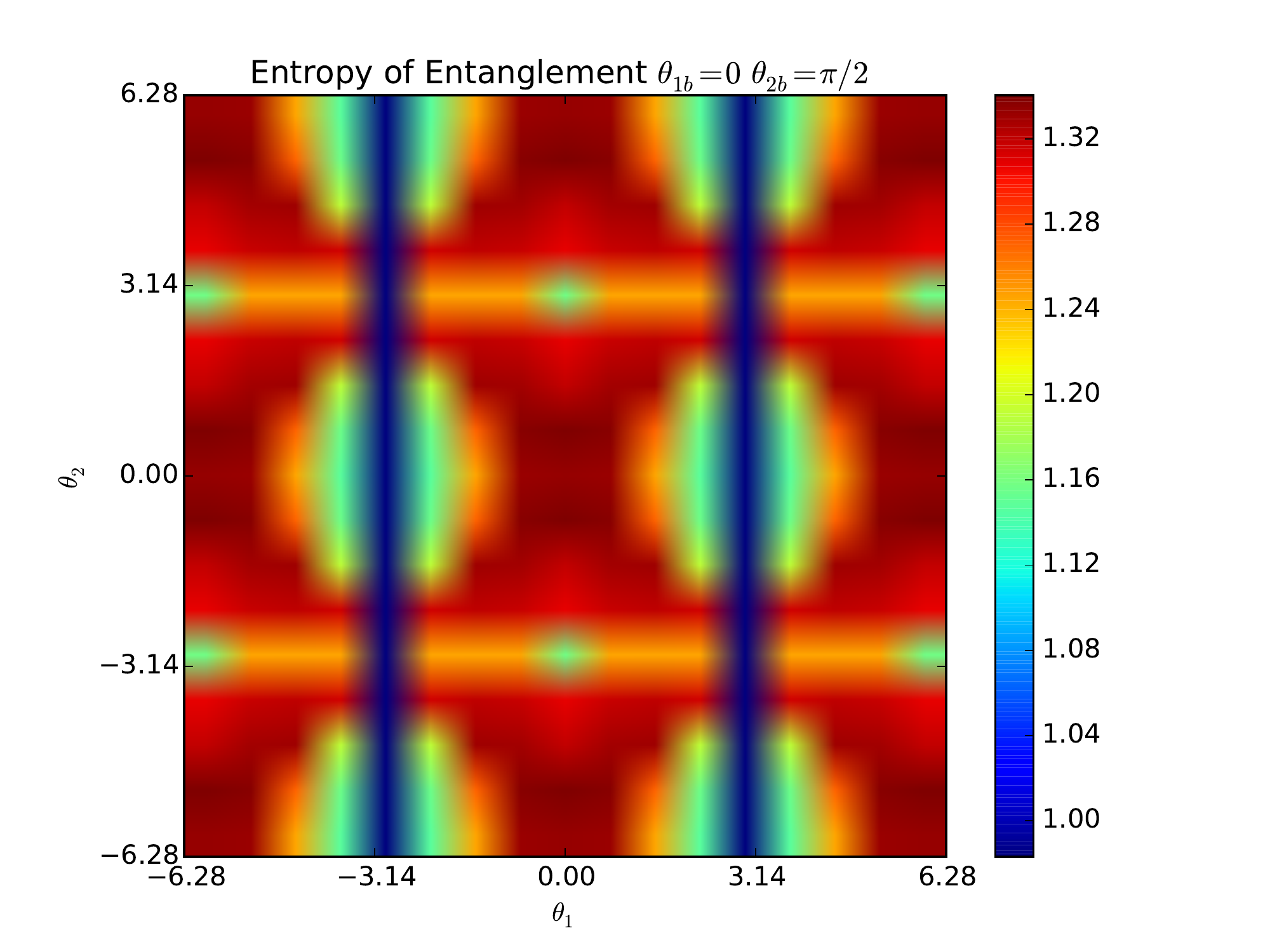}
\caption{TPTPW with $Z_B = 0$ and $\ket{\psi^+_0}$.}\label{fig:5a}
\end{subfigure}\hspace*{\fill}
\begin{subfigure}{0.48\textwidth}
\includegraphics[width=\linewidth]{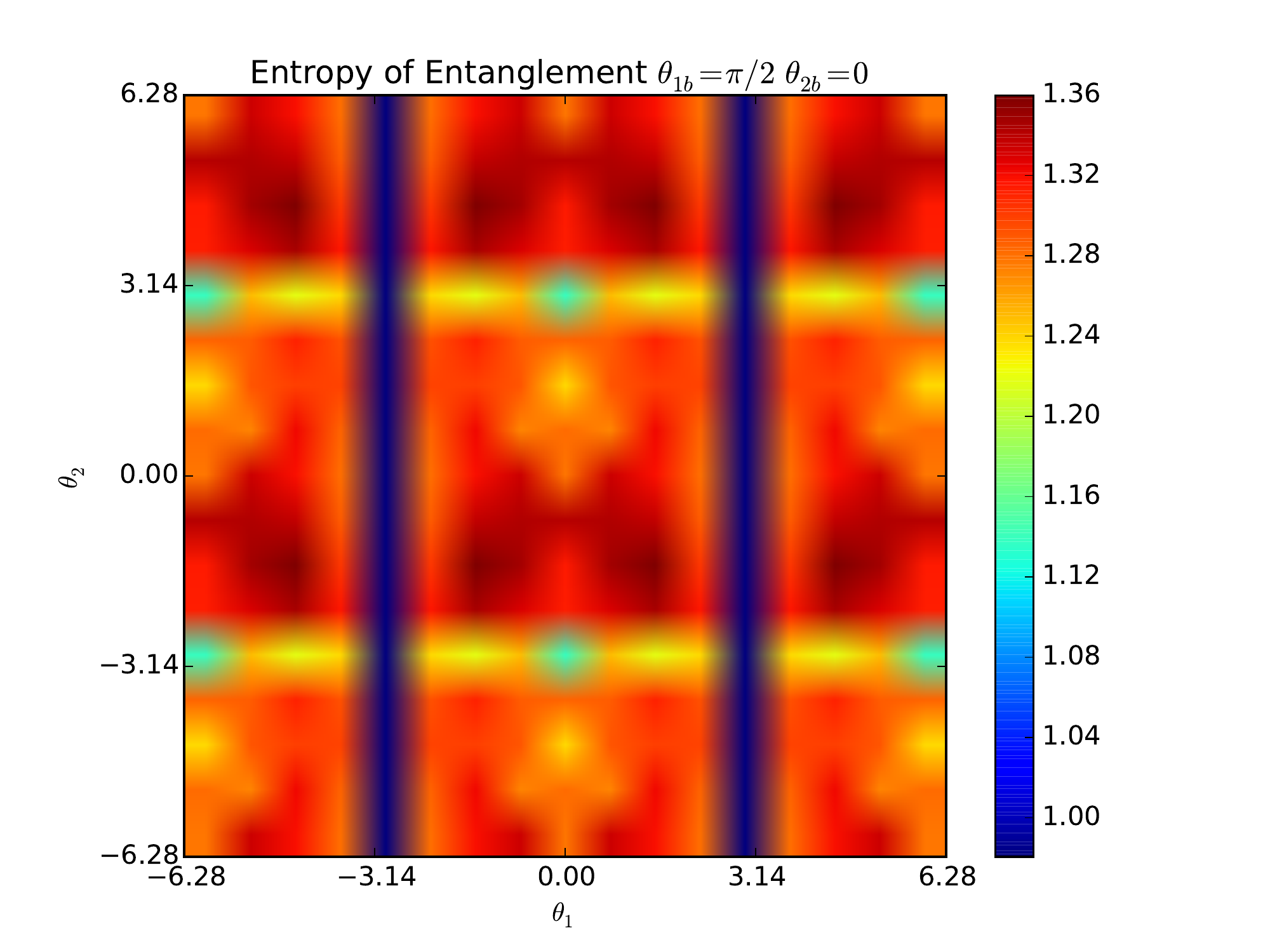}
\caption{TPTPW with $Z_B = 1$ and $\ket{\psi^+_0}$.}\label{fig:5b}
\end{subfigure}

\medskip
\begin{subfigure}{0.48\textwidth}
\includegraphics[width=\linewidth]{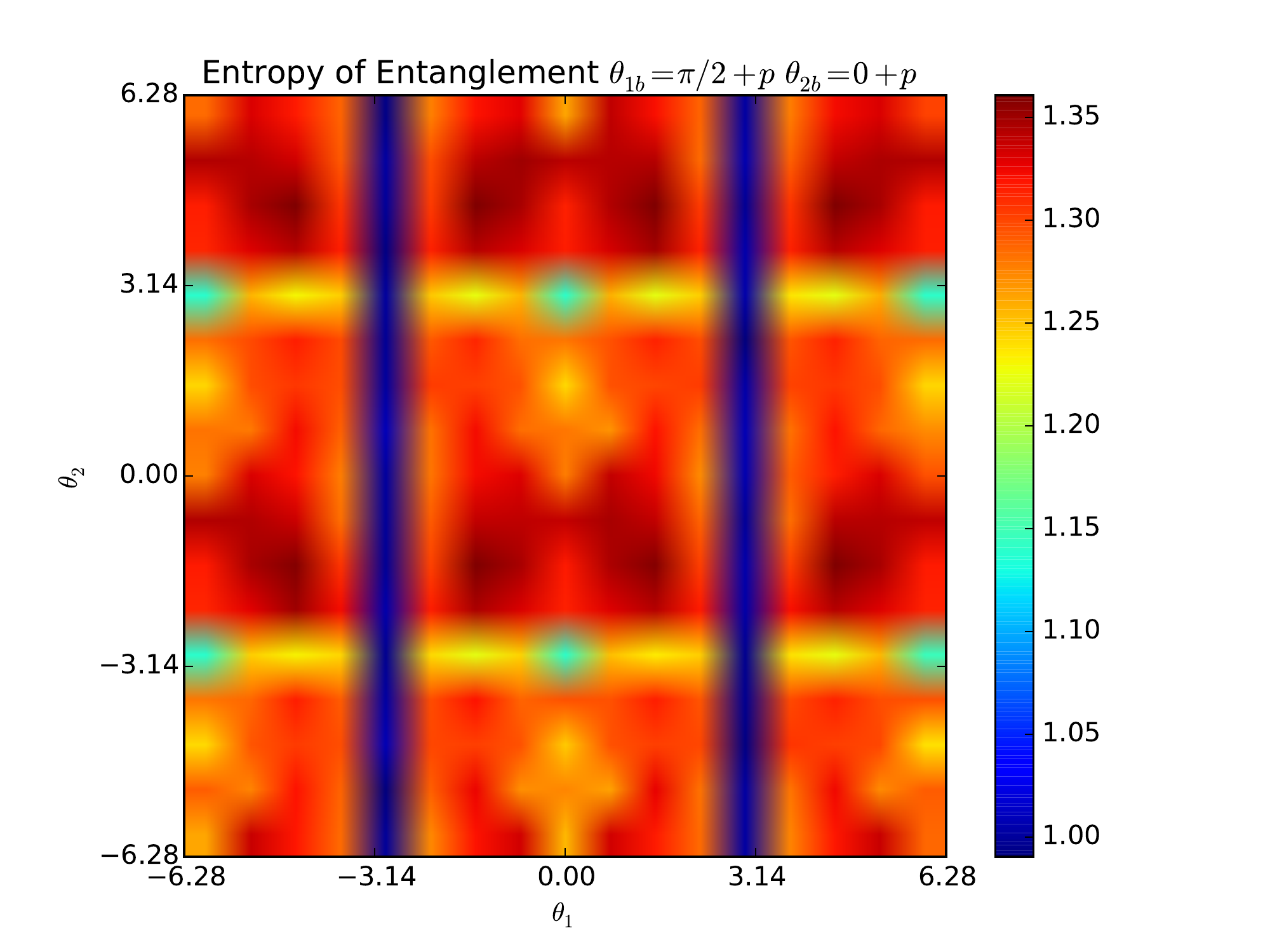}
\caption{TPTPW with $Z_B = 1$ and $\ket{\psi^+_0}$ and weak disorder.}\label{fig:5c}
\end{subfigure}\hspace*{\fill}
\begin{subfigure}{0.48\textwidth}
\includegraphics[width=\linewidth]{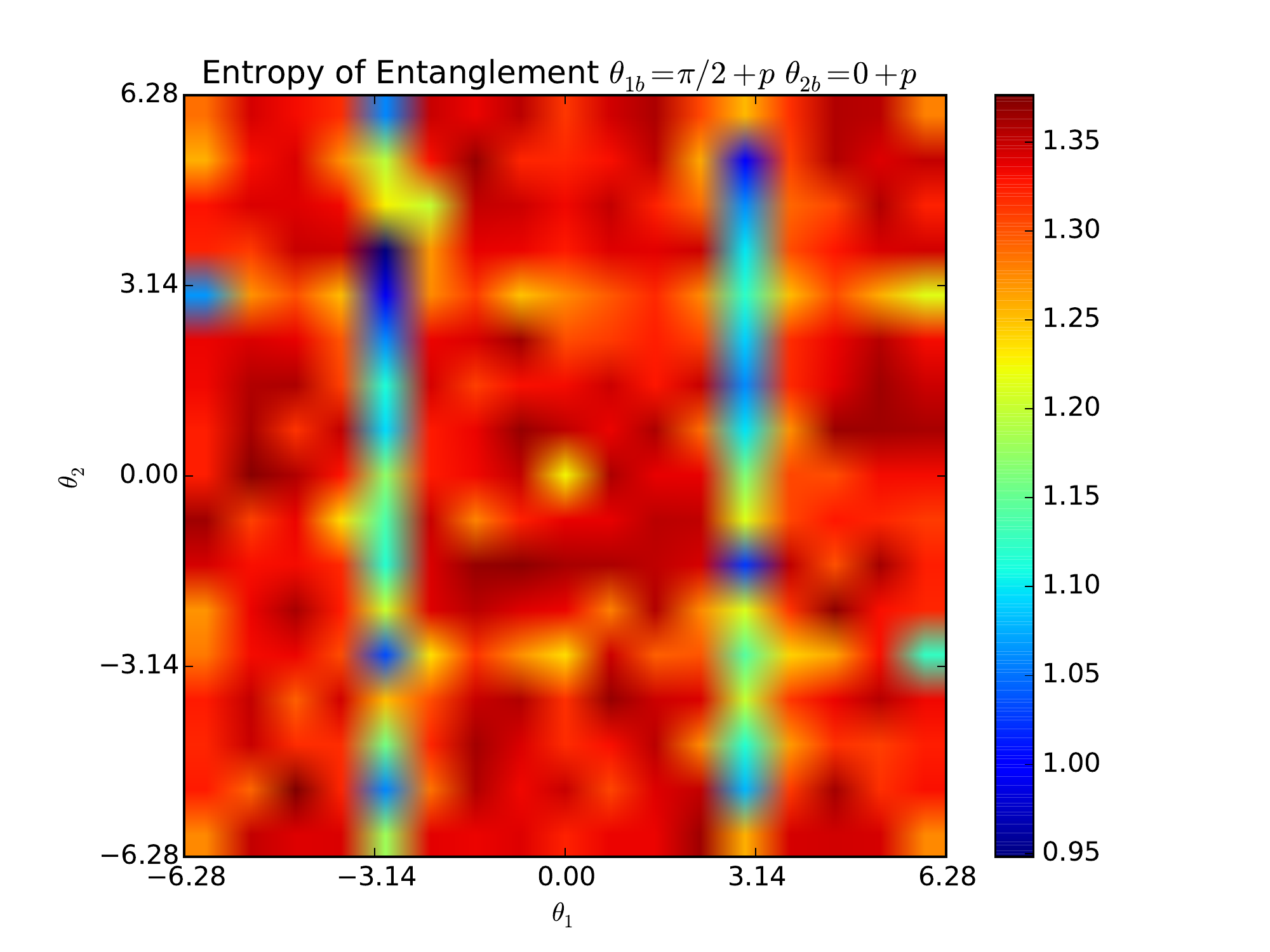}
\caption{TPTPW with $Z_B = 1$ and $\ket{\psi^+_0}$ and strong disorder.}\label{fig:5d}
\end{subfigure}

\medskip
\begin{subfigure}{0.48\textwidth}
\includegraphics[width=\linewidth]{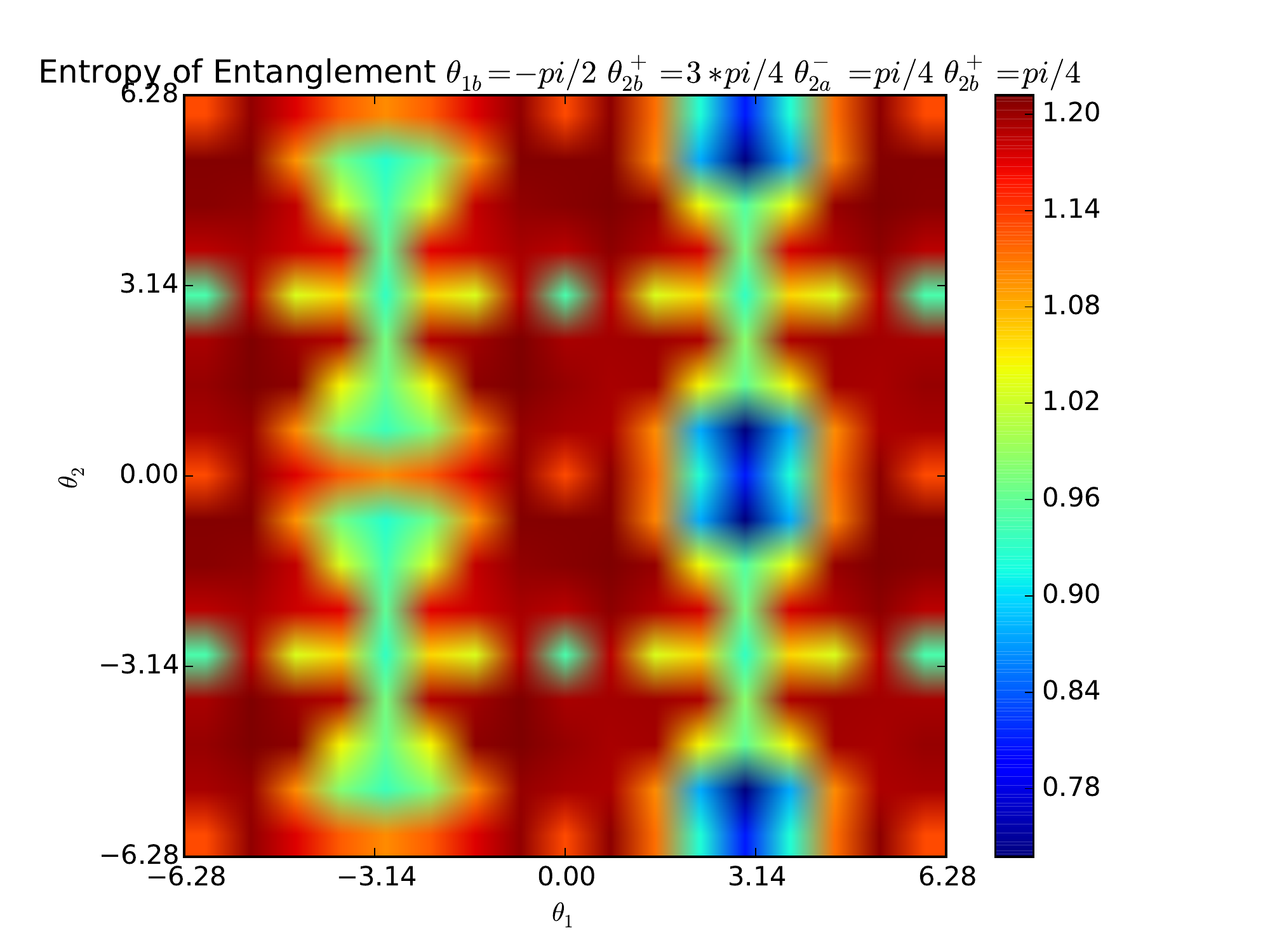}
\caption{TPTBW as function of $\theta_{1A}$ and $\theta^+_{2A}$ for $\ket{\psi^+_0}$.}\label{fig:5e}
\end{subfigure}\hspace*{\fill}
\begin{subfigure}{0.48\textwidth}
\includegraphics[width=\linewidth]{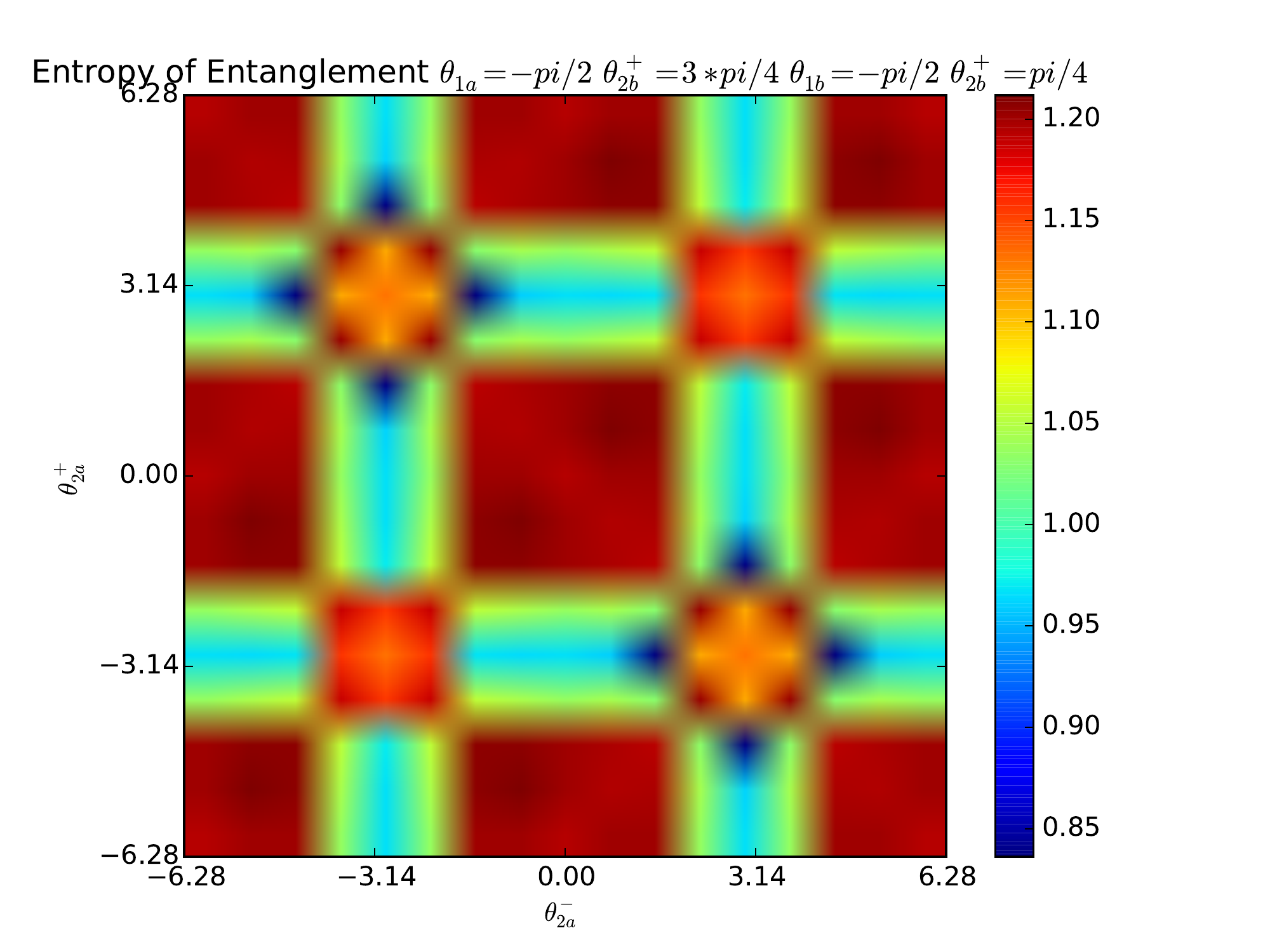}
\caption{TPTBW as function of $\theta^-_{2A}$ and $\theta^+_{2A}$ for $\ket{\psi^+_0}$.}\label{fig:5f}
\end{subfigure}

\caption{$S(\rho_c)$ for TPTPW and TPTBW.}\label{fig:5}
\end{figure*}

\section{Discussion}

An example of $S(\rho_c)$ for TPTPW and TPTBW as function of the number of time steps is shown in Figures ~\ref{fig:3a} and ~\ref{fig:3b}. For the TPTPW, the both the clean and disordered cases trend to the same value, while for the TPTBW case, the mean value of $S(\rho_c)$ of the disordered case is greater than that of the clean case. For both the TPTPW and TPTBW, the fluctuations in $S(\rho_c)$ are greater in the disordered case than the clean case. These fluctuations warrant further investigation.

From these preliminary calculations, there does not appear to be a strong dependence of the topology on $S(\rho_c)$. In other words, differences between the topological phase experienced by particle A and the topological phase experienced by particle B do not manifest in a unique dependence in the $S(\rho_c)$. This is shown in Figures ~\ref{fig:5a} and ~\ref{fig:5b}. As the the topological phase of particle A is varied from $Z_A = 1$ (i.e. $\theta_1 =-\pi/2, \theta_2 = \pi/4 $) to $Z_A = 0$  (i.e. $\theta_1 = -\pi/2, \theta_2 = 3\pi/4 $) , the same $S(\rho_c)$ can be obtained. However, once the walk ($U_{ss,AB}$) is within a set of topological phases, the $S(\rho_c)$ will vary as a function of  $\theta_1, \theta_2$.

Disorder, in its current implementation, has an impact on $S(\rho_c)$ for the TPTPW. In the strong disorder case, $Z_B$ is undefined and the sharp contrasts as a function of $\theta_{1A}, \theta_{2A}$ begin to fade, as seen in Figures ~\ref{fig:5c} and ~\ref{fig:5d}. However, the absolute range of $S(\rho_c)$ for disordered cases remains nearly identical to the range of $S(\rho_c)$ for clean cases. A possible explanation for this weak dependence of the range of $S(\rho_c)$ on disorder could be attributed to the implementation of disorder in the present calculation. Since disorder is introduced trough the spin degree of freedom, it is possible that this will not change the entanglement with position degree of freedom. By introducing disorder in the position, with a percolation model for example as studied in \cite{rigovacca2015two}, this hypothesis can be tested.

For the TPTBW, the behavior of $S(\rho_c)$, is less symmetrical than that of the TPTPW (see Figures ~\ref{fig:5e} and ~\ref{fig:5f}). This can be understood as consequence of, for these specific cases, $\theta_2$ being a function of $x$. A similar asymmetry can be achieved by making $\theta_1$ a function of $x$. It is interesting to see the contrast of low and high $S(\rho_c)$ within a small window of $\theta^{-}_{2A}$ and $\theta^{+}_{2A}$. 

The overall $S(\rho_c)$ can be greater in a TPTPW (approximately 1.3) than a TPTBW (approximately 1.2), thus if entanglement is to be used as a resource, a TPTPW might prove to be worthy of consideration.

On the other hand, disorder clearly impacts the two particle spatial distribution, as can be seen in Figures ~\ref{fig:4b} and ~\ref{fig:4c}. For the TPTPW, disorder localizes the probability about the origin. This can be understood, again, as a consequence of the implementation of disorder, which randomizes the coin operator at each site for each time step, which when coupled with the split translation operator will lead to a position of 0 being most probable. In other words, the classical probability the particle (A and B) moving left or right is reduced to zero. 

For the TPTBW, a boundary state is observed, associated with the non-vanishing probability at the origin, where the two distinct topological phases meet. In the strong disorder case, this boundary mode is wiped out due to the weakening of the distinction between the two topological phases, so the boundary state is robust only to a point. Rakovszky and Asboth have shown that for disorder strengths with a distribution of  $[\theta-\pi,\theta+\pi]$, the topological invariant reduces to zero \cite{PhysRevA.92.052311}, consistent with that is observed in this work.

While not shown in the chosen figures, the effect of initial state on $S(\rho_c)$ and the two particle spatial distribution deserve a remark. For the separable state (see Appendix), $S(\rho_c)$ is on average reduced while the pattern as a function of $\theta_1, \theta_2$ is similar to that of Figures ~\ref{fig:5a} and ~\ref{fig:5b}, while for $\ket{\psi^+_0}$ and $\ket{\psi^-_0}$, the $S(\rho_c)$ exhibit only minor differences. The chosen initial state does not change the earlier remarks with regards to disorder and $S(\rho_c)$. As noted in the simple Hadamard walk, the two particle spatial distribution depends on the initial state. However, this dependence is washed out in the strong disorder case. 

\section{Conclusions}

The lack of clear dependence of $S(\rho_c)$ on the underlying topology of the quantum walk operator is somewhat surprising since systems in which two or more topological phases are present exhibit, amongst other phenomena, boundary states and thus it is reasonable to expect there to be different dynamics. So perhaps, entropy of entanglement is not a clear metric of the effect of distinct topological phases on quantum walkers. Naturally, it would be beneficial to understand the reason for this apparent independence. One interesting path is to perform a closer examination of the fluctuations in the entropy of entanglement, which as seen in Figures ~\ref{fig:3a} and ~\ref{fig:3b}, are influenced by the presence of disorder. Going beyond the measure of entropy of entanglement, we could calculate such quantities as discord, which measures the difference between the quantum and classical mutual information. Ultimately, a metric that is sensitive to the topology is desired. Another potential avenue would be to examine the impact of introducing an interaction between the two particles \cite{berry2011two}. Finally, we can ask if quantum walks with topological phases have an advantage over traditional quantum walks when applied to such problems as quantum search and graph isomorphism determination.

\bibliography{biblio}
\newpage
\section{Appendix}

\subsection{Code}
All the code used in the calculations for this work is available at \\
https://github.com/schuberm/qrw/blob/master/qrw3.ipynb.

\subsection{Supplementary Figures}

\begin{figure*}[ht] 
\begin{subfigure}{0.48\textwidth}
\includegraphics[width=\linewidth]{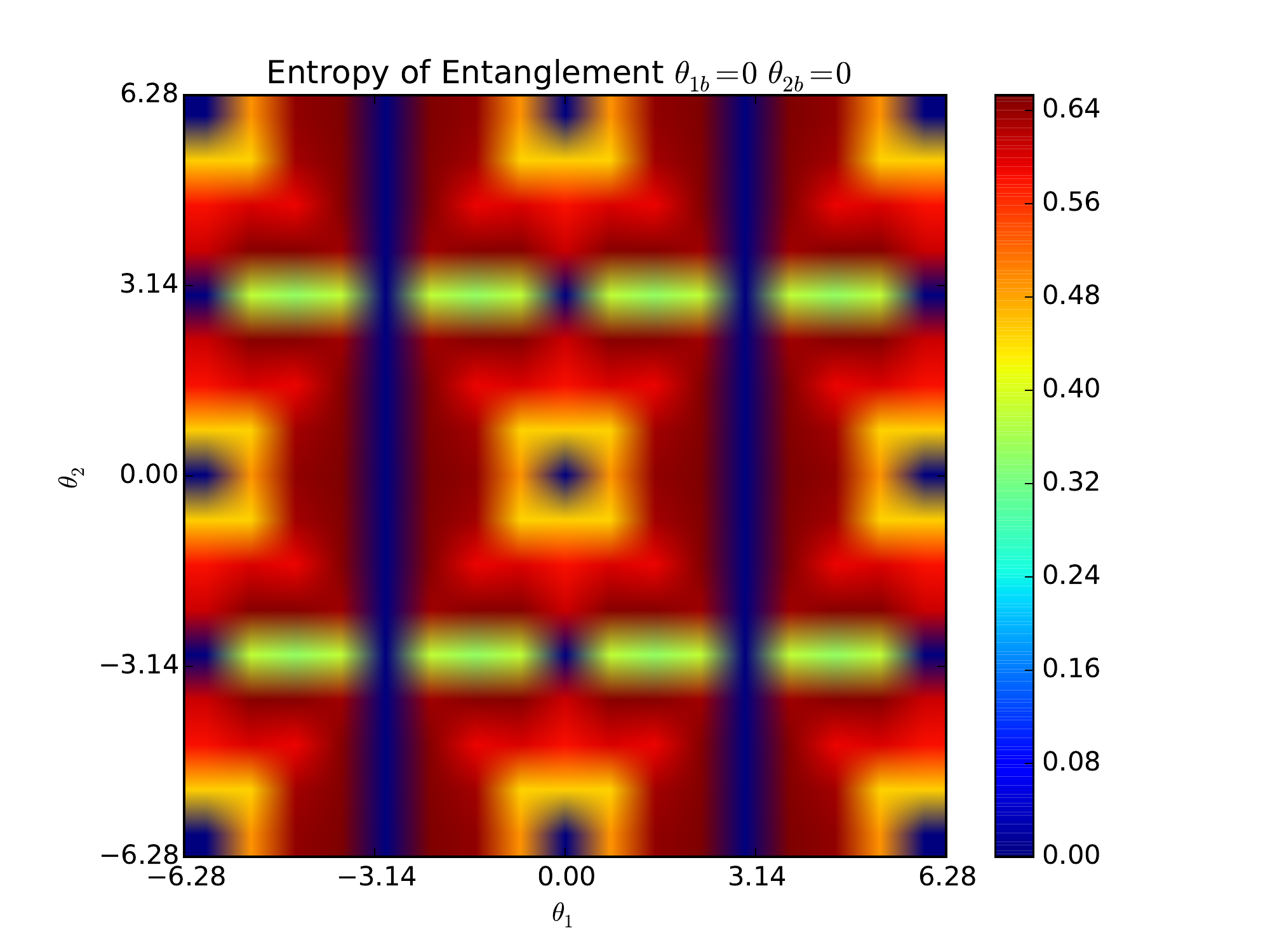}
\caption{TPTPW with $\ket{\psi_0}$ and $Z_B = 1$} \label{fig:6a}
\end{subfigure}\hspace*{\fill}
\begin{subfigure}{0.48\textwidth}
\includegraphics[width=\linewidth]{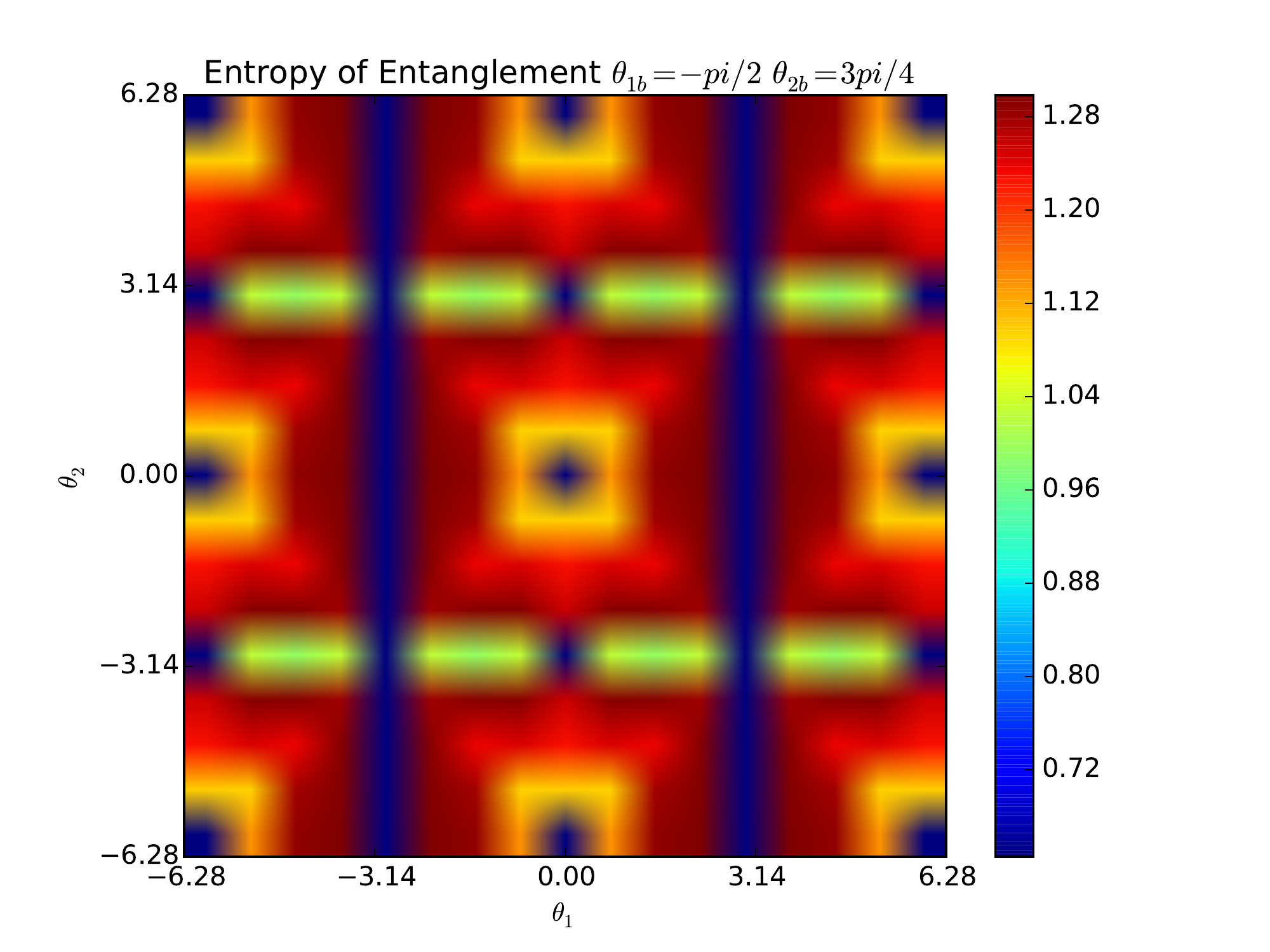}
\caption{TPTPW with $\ket{\psi_0}$ and $Z_B = 0$} \label{fig:6b}
\end{subfigure}

\medskip
\begin{subfigure}{0.48\textwidth}
\includegraphics[width=\linewidth]{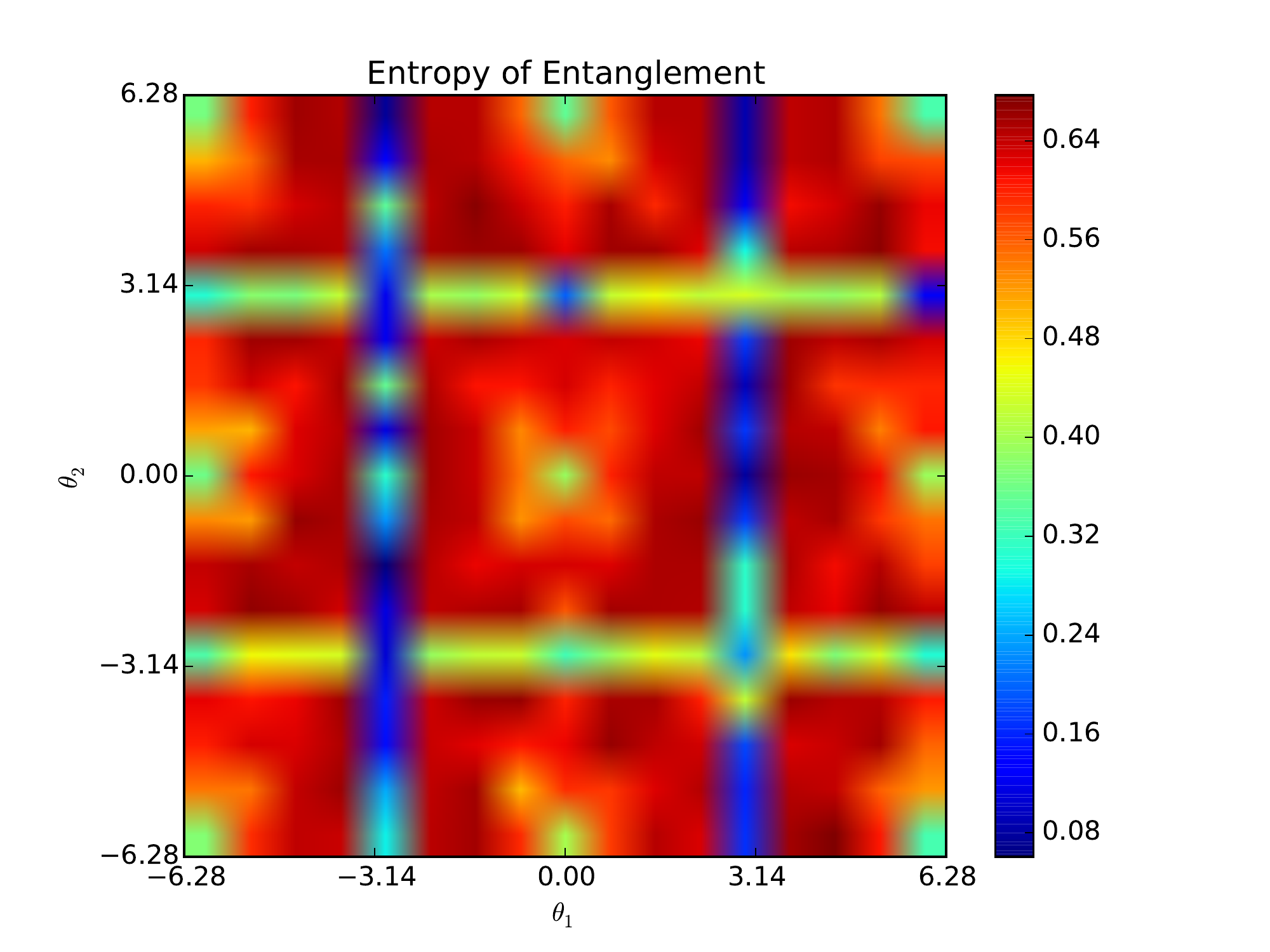}
\caption{Entropy of entanglement for single particle with weak disorder.} \label{fig:6c}
\end{subfigure}\hspace*{\fill}
\begin{subfigure}{0.48\textwidth}
\includegraphics[width=\linewidth]{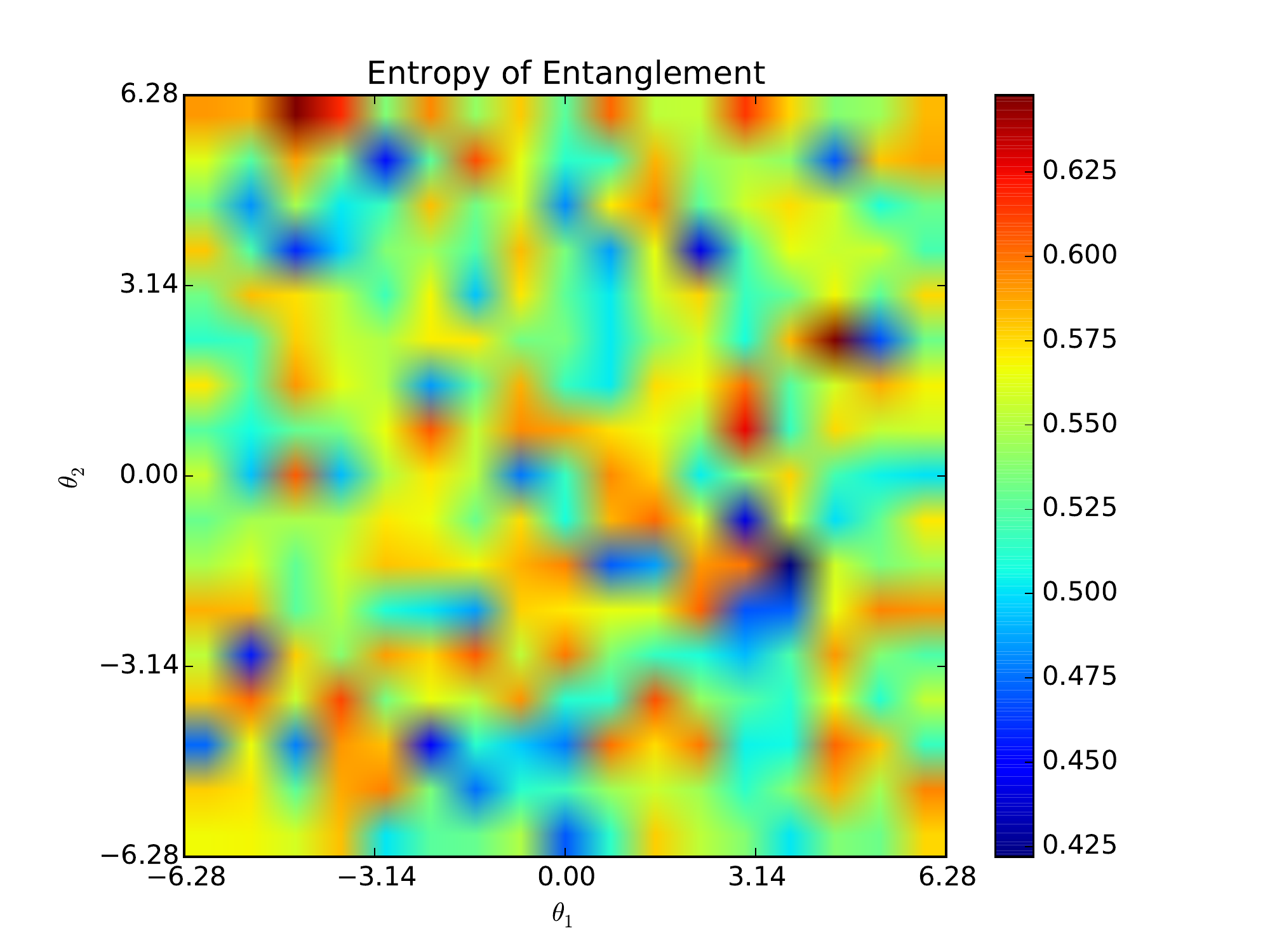}
\caption{Entropy of entanglement for single particle with strong disorder.} \label{fig:6d}
\end{subfigure}

\end{figure*}

\end{document}